\documentclass[10pt, a4paper, twocolumn, twoside]{IEEEtran}

\usepackage{cite}
\usepackage{enumerate}
\usepackage{graphicx}
\usepackage[cmex10]{amsmath} 
\usepackage{amssymb}
\usepackage{xspace}
\usepackage{subfigure}
\usepackage[lined,linesnumbered,ruled,commentsnumbered]{algorithm2e} 
\usepackage{colonequals}
\usepackage[mathscr]{euscript}
\usepackage{fixltx2e}    
\usepackage{amsthm}
\usepackage{mathrsfs}

\usepackage{epsfig}
\usepackage{epstopdf}

\usepackage{tikz}
\usepackage{pgfplots}

\usetikzlibrary{arrows,shapes,backgrounds,plotmarks,positioning}

\pgfplotsset{compat=newest}                         
\pgfplotsset{plot coordinates/math parser=false}
\newlength\figureheight
\newlength\figurewidth

\newtheorem{theorem}{Theorem}[section]
\newtheorem{lemma}[theorem]{Lemma}

\newtheorem{example}[theorem]{Example}

\newcommand{\Has}[1]{W_{#1}}
\newcommand{\rv}{\mathbf{r}}         
\newcommand{\wv}{\mathbf{w}}         

\newcommand{\Fu}[1]{f_{#1}}
\newcommand{\FuHash}[1]{f_{#1}^{\#}}
\newcommand{\FuHashHat}[1]{\hat{f}_{#1}^{\#}}

\newcommand{\Real}{\mathbb{R}}
\newcommand{\RealP}{\mathbb{R}_{+}}    
\newcommand{\Z}{\mathbb{Z}}            
\newcommand{\F}{\mathbb{F}}            

\newcommand{\SFM}{\text{SFM}}

\newcommand{\op}{\text}
\newcommand{\RZ}[1]{\mathsf{Z}_{#1}}
\newcommand{\RW}[1]{\mathsf{W}_{#1}}
\newcommand{\RRCO}{\mathscr{R}_{\op{CO}}}
\newcommand{\RRACO}{\mathscr{R}_{\op{ACO}}}
\newcommand{\RRNCO}{\mathscr{R}_{\op{NCO}}}
\newcommand{\RACO}{R_{\op{ACO}}}
\newcommand{\RNCO}{R_{\op{NCO}}}
\newcommand{\Card}[1]{|#1|}
\newcommand{\Set}[1]{\{#1\}}

\newcommand{\Pat}{\mathcal{P}}

\newcommand{\Y}{\mathcal{Y}}

\newcommand{\X}{\mathcal{X}}

 {\begin{list}{}%
         {\setlength{\leftmargin}{#1}}%
         \item[]%
 }
 {\end{list}}

\begin{document}

\title{A Faster Algorithm for Asymptotic Communication for Omniscience}

\author{Ni~Ding\IEEEauthorrefmark{1}, Chung~Chan\IEEEauthorrefmark{3}, Qiaoqiao~Zhou\IEEEauthorrefmark{3}, Rodney~A.~Kennedy\IEEEauthorrefmark{1} and Parastoo~Sadeghi\IEEEauthorrefmark{1}

\thanks{\IEEEauthorblockA{\IEEEauthorrefmark{1}Ni Ding, Rodney A. Kennedy and Parastoo Sadeghi are with the Research School of Engineering, the Australian National University (email: $\{$ni.ding, rodney.kennedy, parastoo.sadeghi$\}$@anu.edu.au).} Part of the work of Ni Ding (email: ni.ding@inc.cuhk.edu.hk) has been done when she was a junior research assistant at the Institute of Network Coding, Chinese University of Hong Kong, from Nov 23, 2015 to Feb 6, 2016.}
\thanks{\IEEEauthorblockA{\IEEEauthorrefmark{3}Chung Chan (email: cchan@inc.cuhk.edu.hk) and Qiaoqiao Zhou (email: zq115@ie.cuhk.edu.hk) are with the Institute of Network Coding, Chinese University of Hong Kong. }}
}


\maketitle

\begin{abstract}
    We propose a modified decomposition algorithm (MDA) to solve the asymptotic communication for omniscience (CO) problem where the communication rates could be real or fractional. By starting with a lower estimation of the minimum sum-rate, the MDA algorithm iteratively updates the estimation by the optimizer of a Dilworth truncation problem until the minimum is reached with a corresponding optimal rate vector. We also propose a fusion method implementation of the coordinate-wise saturation capacity algorithm (CoordSatCapFus) for solving the Dilworth truncation problem, where the minimization is done over a fused user set with a cardinality smaller than the original one. We show that the MDA algorithm is less complex than the existing ones. In addition, we show that the non-asymptotic CO problem, where the communication rates are integral, can be solved by one more call of the CoordSatCapfus algorithm. By choosing a proper linear ordering of the user indices in the MDA algorithm, the optimal rate vector is also the one with the minimum weighted sum-rate.
\end{abstract}


\section{introduction}

Communication for omniscience (CO) is a problem proposed in \cite{Csiszar2004}. It is assumed that there is a group of users in the system and each of them observes a component of a discrete memoryless multiple source in private. The users can exchange their information over lossless broadcast channels so as to attain \textit{omniscience}, the state that each user obtains the total information in the entire multiple source in the system. The CO problem in \cite{Csiszar2004} is based on an asymptotic source model, where the communication rates could be real or fractional. Meanwhile, coded cooperative data exchange (CCDE) problem proposed in \cite{Roua2010} can be considered a non-asymptotic CO problem where the communication rates are required to be integral. By incorporating the idea of packet-splitting, the CCDE problem can be easily extended to an asymptotic setting.

Determining a rate vector that achieves omniscience with the minimum sum-rate is a fundamental problem in CO. Although the non-asymptotic CO problem has been frequently studied in the literature, there still does not exist an efficient algorithm for the asymptotic setting. The reasons are explained as follows. The submodularity of the CO problem has been shown in \cite{ChanMMI,ChanSuccessive,ChanSuccessiveIT,MiloIT2015,CourtIT2014,Ding2015ICT,Ding2015ISIT}. By designating a sum-rate, a submodular function minimization (SFM) algorithm can check whether the sum-rate is achievable for CO and/or return an achievable rate vector with the given sum-rate. Since the SFM algorithm completes in strongly polynomial time, the remaining problem is how to adapt the sum-rate to the minimum. This problem is not difficult for non-asymptotic setting since every adaptation should be integral. For example, the authors in \cite{MiloIT2015,CourtIT2014} proposed efficient adaptation algorithms for non-asymptotic CO problem, the complexity of which only grows logarithmically in the total amount of information in the system.

However, when considering the asymptotic setting, it is not clear how to choose the step size in each adaptation (Improper step sizes may result in an infinite loop). More specifically, even if we know that a sum-rate is over/below the optimum, it is not sure how much we should decrease/increase from the current estimation. On the other hand, the authors in \cite{MiloDivConq2011} proposed a divide-and-conquer (DV) algorithm for the asymptotic setting by repetitively running a decomposition algorithm (DA) in \cite{MinAveCost}. The idea is to first find the fundamental partition \cite{ChanMMI}, the one corresponds to the minimum sum-rate, and then iteratively break each non-singleton element into singletons so that each tuple in the optimal rate vector is determined. However, the DA algorithm is able to not only determine the fundamental partition but also return an optimal rate vector, which we will explain in this paper. Therefore, those further divisions of the fundamental partition in the DV algorithm are not necessary.

In this paper, we propose a modified decomposition algorithm (MDA) for solving the asymptotic CO problem based on the DA algorithm in \cite{MinAveCost}. The MDA algorithm starts with a lower estimation of the minimum sum-rate. In each iteration, the step size is determined based on the finest/minimum partition of a Dilworth truncation problem. We prove the optimality of the output rate vector and show that the estimation sequence converges monotonically upward to the minimum sum-rate. In addition, we propose a fusion method implementation of the coordinate-wise saturation capacity algorithm (CoordSatCapFus) for solving the Dilworth truncation problem. In the CoordSatCapFus algorithm, the SFM in each iteration is done over a fused user set with a cardinality smaller than the original one. We show that the MDA algorithm can reduce the cubic calls of SFM (in the DV algorithm) to quadratic calls of SFM. We do an experiment to show that the fusion method in the CoordSatCapFus algorithm contributes to a considerable reduction in computation complexity when the number of users grows. We also discuss how to solve the non-asymptotic CO problem by one more run of the CoordSatCapFus algorithm. Finally, we show how to choose a proper linear ordering to solve the minimum weighted sum-rate problem.

\section{System Model}
\label{sec:system}

Let $V$ with $\Card{V}>1$ be the finite set that contains the indices of all users in the system. We call $V$ the \textit{ground set}. Let $\RZ{V}=(\RZ{i}:i\in V)$ be a vector of discrete random variables indexed by $V$. For each $i\in V$, user $i$ can privately observe an $n$-sequence $\RZ{i}^n$ of the random source $\RZ{i}$ that is i.i.d.\ generated according to the joint distribution $P_{\RZ{V}}$. We allow users exchange their sources directly so as to let all users $i\in V$ recover the source sequence $\RZ{V}^n$. We consider both asymptotic and non-asymptotic models. In the asymptotic model, we will characterize the asymptotic behavior as the \emph{block length} $n$ goes to infinity. In non-asymptotic model, the communication rates are required to be integer-valued.

Let $\rv_V=(r_i:i\in V)$ be a rate (vector). We call $\rv_V$ an achievable rate if omniscience is possible by letting users communicate with the rates designated by $\rv_V$. Let $r$ be the function associated with $\rv_V$ such that $r(X)=\sum_{i\in X} r_i,\forall X \subseteq V$ with the convention $r(\emptyset)=0$. For $X,Y \subseteq V$, let $H(\RZ{X})$ be the amount of randomness in $\RZ{X}$ measured by Shannon entropy \cite{YeungITBook}
and $H(\RZ{X}|\RZ{Y})=H(\RZ{X \cup Y})-H(\RZ{Y})$ be the conditional entropy of $\RZ{X}$ given $\RZ{Y}$. In the rest of this paper, we simplify the notation $\RZ{X}$ to $X$. It is shown in \cite{Csiszar2004} that an achievable rate must satisfy the Slepian-Wolf constraints:
\begin{equation} \label{eq:SWConstrs}
    r(X) \geq H(X|V\setminus X), \quad \forall X \subset V.
\end{equation}
The interpretation of the Slepian-Wolf constraint on $X$ is: To achieve CO, the total amount of information sent from user set $X$ should be at least complementary to total amount of information that is missing in user set $V \setminus X$. The set of all achievable rate vectors is
$$ \RRCO(V)=\Set{ \rv_V\in\Real^{|V|} \colon r(X) \geq H(X|V\setminus X),\forall X \subset V }. $$

\subsection{Asymptotic and No-asymptotic Models}
In an asymptotic CO model, the minimum sum-rate can be determined by the following linear programming (LP)
\begin{equation} \label{eq:MinSumRate}
    \RACO(V)=\min\Set{r(V) \colon \rv_V\in \RRCO(V)}
\end{equation}
and the set of all optimal rates is
  $$ \RRACO^*(V)=\Set{\rv_V\in \RRACO(V) \colon r(V)=\RACO(V)}. $$
In a non-asymptotic CO model, $H(X) \in \Z_+$ for all $X \subseteq V$ and the minimum sum-rate can be determined by the integer linear programming (ILP)
$\RNCO(V)=\min\Set{r(V) \colon \rv_V\in \RRCO(V) \cap \Z^{|V|} }$.
The optimal rate set is
$ \RRNCO^*(V)=\Set{\rv_V \in \RRCO(V) \cap \Z^{|V|} \colon r(V)=\RNCO(V)}$.

\subsection{Corresponding CCDE Systems}

CCDE is an example of CO, where the asymptotic model corresponds to the CCDE system that allows packet-splitting, while the non-asymptotic model corresponds to the CCDE system that does not allow packet-splitting. In CCDE, $\RZ{i}$ is the packet set that is obtained by user $i$, where each packet $\Has{j}$ belongs to a field $\F_q$. The users are geographically close to each other so that they can transmit linear combinations of their packet set via lossless wireless channels to help the others recover all packets in $\RZ{V}=\cup_{i\in V} \RZ{i}$. In this problem, the value of $H(X)$ can be obtained by counting the number of packets in $\RZ{X}$, i.e., $H(X)=|\RZ{X}|$ and $H(X|Y)=|\RZ{X \cup Y}|-|\RZ{Y}|$.

\begin{example} \label{ex:main}
Let $V=\{1,\dotsc,5\}$. Each user observes respectively
    \begin{equation}
        \begin{aligned}
            \RZ{1} & = (\RW{a},\RW{c},\RW{e},\RW{f}),   \\
            \RZ{2} & = (\RW{a},\RW{d},\RW{h}),   \\
            \RZ{3} & = (\RW{b},\RW{c},\RW{e},\RW{f},\RW{g},\RW{h}), \\
            \RZ{4} & = (\RW{a},\RW{c},\RW{f},\RW{g},\RW{h}),  \\
            \RZ{5} & = (\RW{b},\RW{d},\RW{f}),
        \end{aligned}  \nonumber
    \end{equation}
where $\RW{j}$ is an independent uniformly distributed random bit. The users exchange their private observations to achieve the omniscience of $\RZ{V}=(\RW{a},\dotsc,\RW{h})$. In this system, $\RACO(V)= \frac{11}{2}$ and $\RNCO(V) = 6$. $\rv_V = (0,\frac{1}{2},2,\frac{5}{2},\frac{1}{2})$ is an optimal rate in $\RRACO^*(V)$ for asymptotic model, while $\rv_V=(0,1,2,3,0)$ is the optimal rate in $\RRNCO^*(V)$ for non-asymptotic model. The method to implement rate $\rv_V = (0,\frac{1}{2},2,\frac{5}{2},\frac{1}{2})$ is to let users divide each packets into two chunks of equal length and transmit according to rate $(0,1,4,5,1)$ with each tuple denotes the number of packet chunks. $(0,\frac{1}{2},2,\frac{5}{2},\frac{1}{2})$ and $\frac{11}{2}$ are the normalized rate and sum-rate, respectively.
\end{example}

\section{Preliminaries}

In this section, we list some existing results derived previously in \cite{ChanMMI,ChanSuccessive,ChanSuccessiveIT,CourtIT2014,MiloIT2015,Ding2015ICT,Ding2015Game,Ding2015ISIT,Ding2015NetCod,FujishigePolyEntropy,Fujishige2005} for CO.

\subsection{Submodularity and Nonemptiness of Base Polyhedron}

It is shown in \cite{FujishigePolyEntropy,Fujishige2005} that the entropy function $H$ is the rank function of a polymatroid, i.e., it is (a) normalized: $H(\emptyset) = 0 $; (b) monotonic: $H(X) \geq H(Y)$ for all $X,Y \subseteq V$ such that $Y \subseteq X$; (c) submodular:
\begin{equation} \label{eq:SubMIneq}
    H(X) + H(Y) \geq H(X \cap Y) + H(X \cup Y)
\end{equation}
for all $X,Y \subseteq V$. For $\alpha\in\RealP$, define the set function $\Fu{\alpha}$ as
$$ \Fu{\alpha}(X)=\begin{cases} H(X|V\setminus X) & X \subset V \\ \alpha & X=V \end{cases}. $$
Let $ \FuHash{\alpha}(X)=\Fu{\alpha}(V)-\Fu{\alpha}(V \setminus X)=\alpha-\Fu{\alpha}(V \setminus X), \forall X \subseteq V$
be the \textit{dual set function} of $\Fu{\alpha}$. It is shown in \cite{ChanMMI,Ding2015ISIT,Ding2015NetCod} that $\FuHash{\alpha}$ is intersecting submodular, i.e., $\FuHash{\alpha}(X) + \FuHash{\alpha}(Y) \geq \FuHash{\alpha}(X \cap Y) + \FuHash{\alpha}(X \cup Y)$ for all $X,Y \subseteq V$ such that $X \cap Y \neq \emptyset$. The polyhedron and base polyhedron of $\FuHash{\alpha}$ are respectively
\begin{equation}
    \begin{aligned}
        & P(\FuHash{\alpha},\leq) = \Set{\rv_V\in\Real^{|V|} \colon r(X) \leq \FuHash{\alpha}(X),\forall X \subseteq V},  \\
        & B(\FuHash{\alpha},\leq) = \Set{\rv_V \in P(\FuHash{\alpha},\leq) \colon r(V) = \FuHash{\alpha}(V)}.  \\
    \end{aligned}  \nonumber
\end{equation}
It is shown in \cite{Ding2015ICT,Ding2015ISIT,Ding2015Game} that $B(\FuHash{\alpha},\leq) = \Set{ \rv_V \in \RRCO(V) \colon r(V) = \alpha}$, i.e., $B(\FuHash{\alpha},\leq)$ denotes the set of all achievable rates with sum-rate equal to $\alpha$, and $B(\FuHash{\alpha},\leq) \neq \emptyset$ if and only if $\alpha \geq \RACO(V)$. In addition, $B(\FuHash{\RACO(V)},\leq) = \RRACO^*(V)$ and $B(\FuHash{\RNCO(V)},\leq) \cap \Z^{|V|} = \RRNCO^*(V)$.

Denote $\Pi(V)$ the set that contains all possible partitions of $V$ and $\Pi'(V)=\Pi(V)\setminus\{V\}$. For $\Pat \in \Pi(V)$, let $\FuHash{\alpha}[\Pat] = \sum_{X \in \Pat} \FuHash{\alpha}(X)$. The Dilworth truncation of $\FuHash{\alpha}$ is \cite{Dilworth1944}
    \begin{equation} \label{eq:Dilworth}
            \FuHashHat{\alpha}(X) = \min_{\Pat\in \Pi(X)} \FuHash{\alpha}[\Pat] , \quad \forall X \subseteq V.
    \end{equation}
If $\alpha \geq \RACO(V)$, $\FuHashHat{\alpha}$ is submodular with $\FuHashHat{\alpha}(V)=\alpha$ and $B(\FuHashHat{\alpha},\leq)=B(\FuHash{\alpha},\leq)$ \cite[Lemma IV.7]{Ding2015Game}.

\subsection{Minimum Sum-rate and Fundamental Partition}

The authors in \cite{Ding2015ISIT,Ding2015Game} show that
\begin{equation} \label{eq:RACO}
    \RACO(V) = \max_{\Pat \in \Pi'(V)} \sum_{X \in \Pat} \frac{ H(V) - H(X) }{|\Pat|-1}
\end{equation}
and $\RNCO(V) = \lceil \RACO(V) \rceil$. Meanwhile, in the studies on secrecy capacity in \cite{ChanMMI,ChanSuccessive,ChanSuccessiveIT}, it is shown that maximum secrecy capacity in $V$ equals to the multivariate mutual information (MMI) $I(V)$, which has a dual relationship with $\RACO(V)$: $\RACO(V)=H(V)-I(V)$, and the finest/minimal maximizer of \eqref{eq:RACO} is called the \textit{fundamental partition} and denoted by $\Pat^*$.

        \begin{algorithm} [t]
	       \label{algo:MDA}
	       \small
	       \SetAlgoLined
	       \SetKwInOut{Input}{input}\SetKwInOut{Output}{output}
	       \SetKwFor{For}{for}{do}{endfor}
            \SetKwRepeat{Repeat}{repeat}{until}
            \SetKwIF{If}{ElseIf}{Else}{if}{then}{else if}{else}{endif}
	       \BlankLine
           \Input{the ground set $V$, an oracle that returns the value of $H(X)$ for a given $X \subseteq V$ and a linear ordering $\Phi = (\phi_1,\dotsc,\phi_{|V|})$}
	       \Output{$\rv_V$ which is a rate vector in the base polyhedron $B(\FuHashHat{\RACO(V)},\leq)$, $\Pat^*$ which is the fundamental partition and $\alpha$ which equals to $\RACO(V)$ }
	       \BlankLine
            initiate $\Pat \leftarrow \{\Set{i} \colon i\in V\}$ and $\alpha \leftarrow \sum_{X \in \Pat}\frac{ H(V) - H(X) }{ |\Pat| - 1 } $ \;
            $(\rv_V,\Pat^*) \leftarrow \text{CoordSatCapFus}(V,H,\alpha,\Phi)$ \;
            \While{$\Pat^* \neq \Pat$}{
                update $\Pat \leftarrow \Pat^*$ and $\alpha \leftarrow \sum_{X \in \Pat^*}\frac{ H(V) - H(X) }{ |\Pat^*| - 1 } $\;
                $(\rv_V,\Pat^*) \leftarrow \text{CoordSatCapFus}(V,H,\alpha,\Phi)$ \;
            }
            return $\rv_V$, $\Pat^*$ and $\alpha$\;
	   \caption{Modified Decomposition Algorithm (MDA) }
	   \end{algorithm}

\section{Algorithm} \label{sec:algo}

In this section, we propose a MDA algorithm, the modified version of the DA algorithm in \cite{MinAveCost}, in Algorithm 1 for solving the asymptotic CO problem and show how to extend it to solve the non-asymptotic one. The MDA algorithm starts with $\alpha$, a lower estimation of $\RACO(V)$, and iteratively updates it by the minimal/finest minimizer of the Dilworth truncation problem $\FuHashHat{\alpha} = \min_{\Pat\in \Pi(V)} \FuHash{\alpha}[\Pat]$ until it reaches the optimal one. The finest minimizer of the Dilworth truncation problem and a rate vector in the base polyhedron $B(\FuHashHat{\alpha},\leq)$ are determined by the CoordSatCapFus algorithm in Algorithm 2. The CoordSatCapFus algorithm is a fusion method to implement the coordinate-wise saturation capacity (CoordSatCap) algorithm that is proposed in \cite{Fujishige2005} and adopted in \cite{MinAveCost} for the Dilworth truncation problem. We list the notations in Algorithms 1 and 2 below.

Let $\chi_X$ be the characteristic vector of the subset $X \subseteq V$. We shorten the notation $\chi_{\Set{i}}$ to $\chi_i$ for a singleton subset of $V$. Let $ \Phi = (\phi_1,\dotsc,\phi_{|V|})$ be a linear ordering of $V$. For example, $\Phi = (2,3,1,4)$ is a linear ordering of $V = \Set{1,\dotsc,4}$. In Section~\ref{sec:MinWeightSumRate}, we will show that by choosing a proper linear ordering of $V$ the output rate $\rv_V$ of Algorithm 1 also minimizes a weighted sum-rate objective function. For $U \subseteq \Pat$ where $\Pat$ is some partition in $\Pi(V)$, denote $\tilde{U}=\cup_{X \in U} X$, i.e., $U$ is a fusion of all the subsets in $U$ into one subset of $V$. For example, for $U = \Set{\Set{1,3},\Set{2,4},\Set{5},\Set{6}} \subset \Set{\Set{1,3},\Set{2,4},\Set{5},\Set{6},\Set{7}} \in \Pi({\Set{1,\dotsc,7}})$, we have $\tilde{U} = \Set{1,\dotsc,6}$. By using these notations, we propose the MDA algorithm for the asymptotic CO problem and show that they can be easily extended to solve the non-asymptotic CO problem as follows.

        \begin{algorithm} [t]
	       \label{algo:PolyRank}
	       \small
	       \SetAlgoLined
	       \SetKwInOut{Input}{input}\SetKwInOut{Output}{output}
	       \SetKwFor{For}{for}{do}{endfor}
            \SetKwRepeat{Repeat}{repeat}{until}
            \SetKwIF{If}{ElseIf}{Else}{if}{then}{else if}{else}{endif}
	       \BlankLine
           \Input{the ground set $V$, an oracle that returns the value of $H(X)$ for a given $X \subseteq V$, $\alpha$ which is an estimation of $\RACO(V)$ and a linear ordering $\Phi = ( \phi_1,\dotsc,\phi_{|V|})$}
	       \Output{$\rv_V$ which is a rate vector in $B(\FuHashHat{\alpha},\leq)$ and $\Pat^*$ which is the minimal/finest minimizer of $\min_{\Pat\in \Pi(V)} \FuHash{\alpha}[\Pat]$ }
	       \BlankLine
            $\rv_V \leftarrow (\alpha - H(V)) \chi_V$ \tcp*{$\rv \in P(\FuHash{\alpha},\leq)$ by doing so.}
            initiate $ r_{\phi_1} \leftarrow \FuHash{\alpha}(\Set{\phi_1})$ and $\Pat^* \leftarrow \Set{\Set{\phi_1}}$\;
            \For{$i=2$ \emph{\KwTo} $|V|$}{
                determine the saturation capacity
                $$ \hat{\xi} \leftarrow \min\Set{ \FuHash{\alpha}(\Set{\phi_i} \cup \tilde{U}) - r(\Set{\phi_i} \cup \tilde{U}) \colon U \subseteq \Pat^*} $$
                and the minimal/smallest minimizer $U^*$\;
                $U_{\phi_i}^* \leftarrow U^* \cup \Set{\phi_i}$\;
                $ \rv_V \leftarrow \rv_V + \hat{\xi} \chi_{\phi_i}$\;
                \tcc{ merge/fuse all subsets in $\Pat^*$ that intersect with $\tilde{U}_{\phi_i}^*$ into one subset $\tilde{U}_{\phi_i}^* \cup \tilde{\X}$}
                $\X \leftarrow \Set{X \in \Pat^* \colon X \cap \tilde{U}_{\phi_i}^* \neq \emptyset}$\;
                $\Pat^* \leftarrow (\Pat^* \setminus \X) \cup \Set{ \tilde{U}_{\phi_i}^* \cup \tilde{\X} }$\;
            }
            return $\rv_V$ and $\Pat^*$\;
	   \caption{Coordinate-wise Saturation Capacity Algorithm by Fusion Method (CoordSatCapFus)}
	   \end{algorithm}

\subsection{Asymptotic Model}

The optimality of the MDA algorithm for the asymptotic setting is summarized in the following theorem with the proof in Appendix~\ref{app:theo:main}, where every step in the CoordSatCapFus algorithm is explained.

\begin{theorem} \label{theo:main}
     The MDA algorithm outputs the minimum sum-rate $\RACO(V)$, the fundamental partition $\Pat^*$ and an optimal rate $\rv_V \in \RRACO(V)$. The estimation of $\RACO(V)$, $\alpha$, converges monotonically upward to $\RACO(V)$.
\end{theorem}

\begin{figure}[tbp]
	\centering
    \scalebox{0.7}{
%
%
%
\definecolor{mycolor1}{rgb}{1,0,1}%
\begin{tikzpicture}

\begin{axis}[%
width=4in,
height=2in,
scale only axis,
xmin=0,
xmax=2,
xtick={0,1,2},
xlabel={\Large iteration index},
xmajorgrids,
ymin=4.7,
ymax=6,
ymajorgrids,
ylabel={\Large $\alpha$},
legend style={at={(0.25,0.98)},anchor=north west,draw=black,fill=white,legend cell align=left}
]
\addplot [
color=blue,
solid,
line width=1.5pt,
mark=asterisk,
mark options={solid},
]
table[row sep=crcr]{
0 4.75\\
1 5.5 \\
2 5.5 \\
};
\addlegendentry{\Large estimation sequence of $\RACO(V)$};

\addplot [
color=mycolor1,
dotted,
line width=2pt,
]
table[row sep=crcr]{
0 5.5\\
1 5.5\\
2 5.5\\
};
\addlegendentry{\Large $\RACO(V)$};

\end{axis}
\end{tikzpicture}%
}
	\caption{The estimation sequence of $\RACO(V)$, i.e., the value of $\alpha$ in each iteration, when the MDA algorithm is applied to the system in Example~\ref{ex:AlgorithmACO}.}
	\label{fig:Converge}
\end{figure}
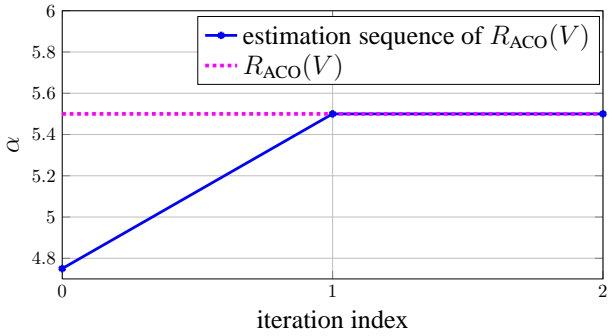

\begin{example} \label{ex:AlgorithmACO}
For the system in Example~\ref{ex:main}, we start the MDA algorithm with singleton partition $\Pat=\Set{\Set{1},\dotsc,\Set{5}}$ and $\alpha = \sum_{i \in V}\frac{ H(V) - H(\Set{i}) }{ |V| - 1 } = \frac{19}{4}$. Let the linear ordering be $\Phi = (4,3,2,5,1)$. By calling the CoordSatCapFus algorithm, we have the following results.

We initiate $\rv_V = (\alpha - H(V)) \chi_V = (-\frac{13}{4},\dotsc,-\frac{13}{4})$ and set $\Pat^*=\Set{\Set{4}}$ and $r_4 = \FuHash{19/4}(\Set{4}) = \frac{7}{4}$ so that $\rv_V = (-\frac{13}{4},-\frac{13}{4},-\frac{13}{4},\frac{7}{4},-\frac{13}{4})$.
\begin{itemize}
    \item For $\phi_2=3$, the values of $\FuHash{\alpha}(\Set{\phi_2} \cup \tilde{U}) - r(\Set{\phi_2} \cup \tilde{U})$ for all $U \subseteq \Pat^* = \Set{\Set{4}}$ are
            $$ \FuHash{19/4}(\Set{3}) - r(\Set{3}) = 6 ,  \FuHash{19/4}(\Set{3,4}) - r(\Set{3,4}) = 21/4.$$
            So, the saturation capacity $\hat{\xi} = 21/4$, the minimal minimizer $U^* = \Set{\Set{4}}$ and $U_4^* = \Set{\Set{3},\Set{4}}$. We update to $r_3 = -\frac{13}{4} + \frac{21}{4} = 2$ so that $\rv_V = (-\frac{13}{4},-\frac{13}{4},2,\frac{7}{4},-\frac{13}{4})$. We have only one element $\Set{4} \in \Pat^*$ such that $\tilde{U}_{4}^* \cap \Set{4} \neq \emptyset$. So, $\X = \Set{\Set{4}}$ and $\tilde{U}_{\phi_i}^* \cup \tilde{\X} = \Set{3,4}$. We update to $\Pat^* = \Set{\Set{3,4}}$.
    \item For $\phi_3=2$, the values of $\FuHash{\alpha}(\Set{\phi_3} \cup \tilde{U}) - r(\Set{\phi_3} \cup \tilde{U})$ for all $U \subseteq \Pat^* = \Set{\Set{3,4}}$ are
            \begin{equation}
                \begin{aligned}
                     &\FuHash{19/4}(\Set{2}) - r(\Set{2}) = 3 , \\ &\FuHash{19/4}(\Set{2,3,4}) - r(\Set{2,3,4}) = 17/4.
                \end{aligned} \nonumber
            \end{equation}
             We have $\hat{\xi} = 3$ and $U_{2}^* = \Set{\Set{2}}$. We update to $\rv_V = (-\frac{13}{4},-\frac{1}{4},2,\frac{7}{4},-\frac{13}{4})$. Since $\tilde{U}_{2}^* \cap X = \emptyset, \forall X \in \Pat$, we have $\X = \emptyset$ and $\Pat^* = \Set{\Set{3,4},\Set{2}}$.
    \item For $\phi_4 = 5$, we have $\hat{\xi}=3$, $U_{5}^*=\Set{\Set{5}}$ and $\X=\emptyset$. We update to $\rv_V = (-\frac{13}{4},-\frac{1}{4},2,\frac{7}{4},-\frac{1}{4})$ and $\Pat^* = \Set{\Set{3,4},\Set{2},\Set{5}}$.
    \item For $\phi_5=1$, we have $\hat{\xi} = \frac{13}{4}$, $U_1^* = \Set{\Set{3,4},\Set{1}}$ and $\X = \Set{\Set{3,4}}$. Therefore, the CoordSatCapFus algorithm terminates with $\rv_V = (0,-\frac{1}{4},2,\frac{7}{4},-\frac{1}{4})$ and $\Pat^*=\Set{\Set{1,3,4},\Set{2},\Set{5}}$.
\end{itemize}
Since $\Pat \neq \Pat^*$, we continue the iteration in the MDA algorithm. In the second iteration, we have $\Pat=\Set{\Set{1,3,4},\Set{2},\Set{5}}$ and $\alpha = \frac{11}{2}$. The CoordSatCapFus algorithm returns $\rv_V = (0,\frac{1}{2},2,\frac{5}{2},\frac{1}{2})$ and $\Pat^* = \Set{\Set{1,3,4},\Set{5},\Set{2}}$. The MDA algorithm terminates since $\Pat = \Pat^*$. One can show that the outputs $\rv_V$, $\Pat^*$ and $\alpha$ are respectively an optimal rate in $\RRACO^*(V)$, the fundamental partition and the minimum sum-rate $\RACO(V)$ for asymptotic model. We plot the value of $\alpha$ in each iteration, or the estimation sequence of $\RACO(V)$, in Fig.~\ref{fig:Converge}. It can be shown that $\alpha$ converges monotonically upward to $\RACO(V)$.
\end{example}

The CoordSatCap algorithm is one of the standard tools for solving the Dilworth truncation problem in the literature, e.g., \cite{MinAveCost}. It is also used in \cite{MiloIT2015,CourtIT2014} to determine an optimal rate vector in $\RRNCO^*(V)$ and/or checking whether a sum-rate $\alpha$ is achievable for non-asymptotic setting.\footnote{If the sum-rate $\alpha$ is not achievable, the rate $\rv_V \in B(\FuHashHat{\alpha},\leq)$ returned by the CoordSatCap algorithm has $r(V)$ strictly less than $\alpha$.} But, in these works, the CoordSatCap algorithm is implemented on the original user set instead of a fused one. For example, in \cite{MiloIT2015,CourtIT2014}, the the saturation capacity $\hat{\xi}$ is determined by the SFM problem
    \begin{equation} \label{eq:SatCapOld}
        \min\Set{ \FuHash{\alpha}(X) - r(X) \mid \phi_i \in X \subseteq V_i},
    \end{equation}
where $V_{i} = \Set{\phi_1,\dotsc,\phi_i}$. Problem \eqref{eq:SatCapOld} can be solved in $O(\SFM(|V_{i-1}|))$ time, where $\SFM(|V|)$ denotes the complexity of an SFM algorithm for a set function defined on $2^{V}$. On the contrary, the corresponding SFM problem
    \begin{equation} \label{eq:SatCapFus}
        \min\Set{ \FuHash{\alpha}(\Set{\phi_i} \cup \tilde{U}) - r(\Set{\phi_i} \cup \tilde{U}) \colon U \subseteq \Pat^*}
    \end{equation}
in step 4 in the CoordSatCapFus algorithm is done over $\Pat^*$, a fused/merged user sets of $V_{i-1}$ that is obtained by steps 8 and 9 in the previous iterations. Here, the objective function in \eqref{eq:SatCapFus} is submodular on $2^{\Pat^*}$. Problem \eqref{eq:SatCapFus} can be solved in $\SFM(|\Pat^*|)$ time. Since $|\Pat^*| \leq |V_2|$, \eqref{eq:SatCapFus} is less complex than \eqref{eq:SatCapOld}. For example, in the first iteration of the MDA algorithm when $\phi_3 = 2$ in Example~\ref{ex:AlgorithmACO}, We have $\Pat^*=\Set{\Set{3,4}}$ and $V_2 = \Set{3,4}$ such that $|\Pat^*|<|V_2|$. Problem~\eqref{eq:SatCapFus} completes in $O(\SFM(1))$ time, while problem \eqref{eq:SatCapOld} completes in $O(\SFM(2))$ time.\footnote{In the case when $|V|=1$, SFM reduces to comparison between two possible sets, empty and ground sets, i.e., it is not necessary to call the SFM algorithm. This example just shows the difference in complexity. } See the experimental results in Section~\ref{sec:Complexity}.

%
%

\subsection{Non-asymptotic Model} \label{sec:NCO}

The algorithms in \cite{MiloIT2015,CourtIT2014} for non-asymptotic CO model can adjust $\alpha$ on the nonnegative integer grid until it finally reaches $\RNCO(V)$, where the CoordSatCap can be replaced by the CoordSatCapFus algorithm which is less complex. See experimental results in Section~\ref{sec:Complexity}.

In fact, the value of $\RNCO(V)$ and an optimal rate in $\RRNCO^*(V)$ can be determined by one more call of the CoordSatCapFus algorithm after solving the asymptotic CO problem. Let $\RACO(V)$ be the asymptotic minimum sum-rate determined by the MDA algorithm. We know automatically $\RNCO(V) = \lceil \RACO(V) \rceil$. By calling the CoordSatCapFus algorithm with input $\alpha = \RNCO(V)$, we can determine the value of an optimal rate in $B(\FuHashHat{\RNCO(V)},\leq) \cap \Z^{|V|} = \RRNCO^*(V)$. The integrality of this optimal vector is shown in Section~\ref{sec:MinWeightSumRate}.

\begin{example} \label{ex:AlgorithmNCO}
Assume that we get $\RACO(V) = \frac{11}{2}$ in Example~\ref{ex:AlgorithmACO}. Then, $\RNCO(V) = \lceil \RACO(V) \rceil = 6$. By calling
$$ (\rv_V,\Pat^*) \leftarrow \text{CoordSatCapFus}(V,H,\RNCO(V),\Phi),$$
we have $\rv_V=(0,1,2,3,0)$ for linear ordering $\Phi = (4,3,2,5,1)$ and $\Pat^* = \Set{\Set{1,2,3,4,5}}$,\footnote{For $\RNCO(V) > \RACO(V) $, the minimizer of $\min_{\Pat\in \Pi(V)} \FuHash{\RNCO(V)}[\Pat]$ is uniquely $\Set{V}$ \cite{Narayanan1991PLP}.} where $\rv_V$ is an optimal rate in $\RRNCO^*(V)$ for non-asymptotic model.
\end{example}

\section{Minimum Weighted Sum-rate Problem} \label{sec:MinWeightSumRate}

Let $\wv_V = (w_i \colon i \in V) \in \RealP^{|V|}$ and $\wv_V^{\intercal} \rv_V = \sum_{i \in V} w_i r_i$. We say that $\Phi = (\phi_1,\dotsc,\phi_{|V|})$ is a linear ordering that is consistent with $\wv_V$ if $w_{\phi_1} \leq w_{\phi_2} \leq \dotsc \leq w_{\phi_{|V|}}$.

\begin{theorem}
    Let $\Phi$ be the linear ordering consistent with $\wv_V$. The optimal rate $\rv_V$ returned by the MDA algorithm for asymptotic model is the minimizer of $\min \Set{ \wv_V^{\intercal} \rv_V \colon \rv_V \in \RRACO^*(V) }$; The optimal rate $\rv_V$ returned by $\text{CoordSatCapFus}(V,H,\lceil \RACO(V) \rceil,\Phi)$ for asymptotic model is the minimizer of $ \min \Set{ \wv_V^{\intercal} \rv_V \colon \rv_V \in \RRNCO^*(V) } $.
\end{theorem}
\begin{IEEEproof}
In the last iteration of the MDA algorithm, we call the CoordSatCapFus algorithm by inputting $\alpha=\RACO(V)$. The Dilworth truncation $\FuHashHat{\RACO(V)}$ is a polymatroid rank function \cite{ChanMMI}. Let $\text{EX}(\FuHashHat{\RACO(V)})$ be the set that contains all extreme points, or vertices, of the base polyhedron $B(\FuHashHat{\RACO(V)},\leq)$. We have the initial point $\rv_V = (\alpha - H(V)) \chi_V \leq \rv_V^\prime, \forall  \rv_V^\prime \in \text{EX}(\FuHashHat{\RACO(V)})$.\footnote{$\rv_V \leq \rv_V^\prime, \forall  \rv_V^\prime \in \text{EX}(f)$ is a tighter condition than $\rv_V \in P(f,\leq)$.} So, the CoordSatCapFus algorithm necessarily returns an extreme point in $B(\FuHashHat{\RACO(V)},\leq)$ which minimizes $\min \Set{ \wv_V^{\intercal} \rv_V \colon \rv_V \in \RRACO^*(V) }$\cite{Fujishige2005}. In the same way, we can prove the claim for the non-asymptotic model. In addition, $\FuHash{\RNCO(V)}$ is integer-valued. So is $\FuHashHat{\RNCO(V)}$. Therefore, all extreme points in $B(\FuHashHat{\RNCO(V)},\leq)$ are integral.
\end{IEEEproof}

For example, one can show that $\rv_V = (0,\frac{1}{2},2,\frac{5}{2},\frac{1}{2})$ in Example~\ref{ex:AlgorithmACO} and $\rv_V=(0,1,2,3,0)$ in Example~\ref{ex:AlgorithmNCO} are the minimum weighted sum-rate vector in $\RRACO^*(V)$ and $\RRNCO^*(V)$, respectively, where the weight $\wv_V$ is the one that linear ordering $\Phi = (4,3,2,5,1)$ is consistent with, e.g., $\wv_V = (4,0.5,0.5,0.3,3.3)$.

Note, any linear ordering is consistent with $\wv_V=(1,\dotsc,1)$, i.e., if the problem is just to determine the minimum sum-rate and an optimal rate vector, the linear ordering can be arbitrarily chosen.

\section{Complexity} \label{sec:Complexity}

The authors in \cite{MiloDivConq2011} proposed a divide-and-conquer (DV) algorithm for the asymptotic CO problem. The idea is to directly apply the DA algorithm in \cite{MinAveCost} to determine the fundamental partition and iteratively break each non-singleton subsets in it into singletons to determine each tuple in the optimal rate. Since the DA algorithm completes in $O(|V|^2\cdot\SFM(|V|))$ time, the complexity of the DV algorithm is upper bounded by $O(|V|^3\cdot\SFM(|V|))$. The complexity of the MDA algorithm is upper bounded by $O(|V|^2\cdot\SFM(|V|))$,\footnote{The complexity of the CoordSatCapFus algorithm based on~\eqref{eq:SatCapFus} in the worst case is the same as the CoordSatCap algorithm based on~\eqref{eq:SatCapOld}. The worst case is when $\Pat^*=\Set{\Set{\phi_1},\dotsc,\Set{\phi_i}}$ for all $i$ in the CoordSatCapFus algorithm. In the DA algorithm in \cite{MinAveCost}, the CoorSatCap algorithm is implemented for solving the Dilworth truncation problem. Therefore the complexity of MDA algorithm is upper bounded by $O(|V|^2\cdot\SFM(|V|))$. } which is lower than the DV algorithm.

\begin{figure}[tbp]
	\centering
    \scalebox{0.7}{
%
%
\begin{tikzpicture}

\begin{axis}[%
width=4in,
height=2.2in,
scale only axis,
xmin=5,
xmax=50,
xmajorgrids,
xlabel={\Large $|V|$},
ymin=0,
ymax=60000,
ymajorgrids,
ylabel={\Large mean size of SFM},
legend style={at={(0.01,0.744916876685679)},anchor=south west,draw=black,fill=white,legend cell align=left}
]
\addplot [
line width=1.5pt,
mark=asterisk,
color=blue,
solid
]
table[row sep=crcr]{
5 16.5\\
6 33.75\\
7 53.55\\
8 109.2\\
9 154.8\\
10 283.5\\
11 407\\
12 501.6\\
13 752.7\\
14 919.1\\
15 1265.25\\
16 1488\\
17 1808.8\\
18 2264.4\\
19 2676.15\\
20 3163.5\\
21 3685.5\\
22 4400.55\\
23 5022.05\\
24 5865\\
25 6675\\
26 7556.25\\
27 8634.6\\
28 9620.1\\
29 10677.8\\
30 12114.75\\
31 13252.5\\
32 14681.6\\
33 16104\\
34 17699.55\\
35 19337.5\\
36 21231\\
37 22977\\
38 25061.95\\
39 26935.35\\
40 29913\\
41 31775\\
42 33751.2\\
43 36526.35\\
44 39353.6\\
45 42322.5\\
46 45177.75\\
47 48536.9\\
48 51549.6\\
49 54919.2\\
50 58616.25\\
};
\addlegendentry{\Large CoordSatCap algorithm based on \eqref{eq:SatCapOld}};

\addplot [
line width=1.5pt,
color=red,
mark=triangle,
solid
]
table[row sep=crcr]{
5 14.25\\
6 30.6\\
7 51.6\\
8 86.75\\
9 114.9\\
10 195.6\\
11 280.35\\
12 331.95\\
13 483.1\\
14 602.55\\
15 728\\
16 805.25\\
17 1093.35\\
18 1294.55\\
19 1577.05\\
20 1734.6\\
21 2082\\
22 2562.55\\
23 3007.05\\
24 3565.7\\
25 4057.9\\
26 4573.05\\
27 5486.35\\
28 5936.65\\
29 6552.2\\
30 7685.15\\
31 8270.05\\
32 8947.6\\
33 10117.35\\
34 11381\\
35 12561\\
36 13626.9\\
37 14850\\
38 15767.6\\
39 17446\\
40 19620.75\\
41 20904.4\\
42 21989.4\\
43 23635.6\\
44 25578.75\\
45 27496.4\\
46 29198.05\\
47 31776.75\\
48 34145.7\\
49 35735.5\\
50 39180\\
};
\addlegendentry{\Large CoordSatCapFus algorithm based on \eqref{eq:SatCapFus}};

\end{axis}
\end{tikzpicture}%
}
	\caption{The size of SFM problem over repetitions in the experiment in Section~\ref{sec:Complexity}, where $H(V)$ is fixed to $50$ and $|V|$ varies from $5$ to $50$.}
	\label{fig:Complexity}
\end{figure}
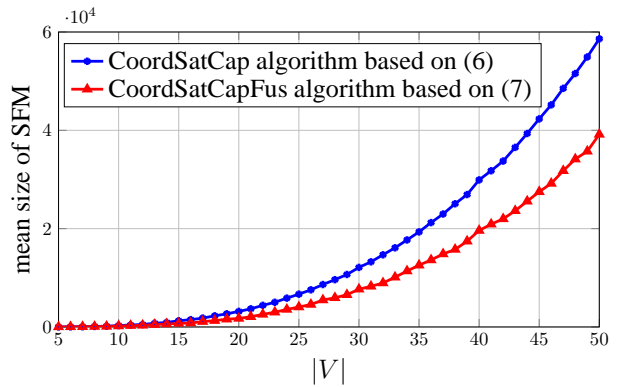

Let $|V|$ be the size of the SFM problem with complexity $\SFM(|V|)$. As aforementioned, although the numbers of calls of SFM algorithm are the same, the size of each SFM problem in the CoordSatCapFus algorithm based on \eqref{eq:SatCapFus} is less than that in the CoordSatCap algorithm based on \eqref{eq:SatCapOld} in general. We do an experiment to compare the complexity of these two algorithms. Let $H(V)$ be fixed to $50$ and $|V|$ vary from $5$ to $30$. For each value of $|V|$, we repeat the procedure for 20 times: (a) randomly generate a CO system; (b) apply the MDA algorithm twice, one calls the CoordSatCapFus algorithm and the other calls the CoordSatCap algorithm. We record overall/summed size of the SFM algorithm in each run of the MDA algorithm and average over the repetitions. The results are shown in Fig.~\ref{fig:Complexity}. It can be seen that by implementing the CoordSatCapFus algorithm, there is a considerable reduction in complexity when the size of user set $|V|$ grows.

\section{Conclusion}

We proposed an MDA algorithm for determining the minimum sum-rate and a corresponding optimal rate for the asymptotic CO problem. The MDA algorithm mainly proposed an idea on how to update the minimum sum-rate estimation: A closer estimation to the optimum could be obtained by the minimal/finest minimizer of a Dilworth truncation problem based on the current estimation. We also proposed a CoordSatCapFus algorithm to solve the Dilworth truncation problem which was less complex than the original CoordSatCap algorithm. We discussed how to extend the MDA algorithm to solve the non-asymptotic problem and how to choose a proper linear ordering of the user set to solve a minimum weighted sum-rate problem.

\appendices

\section{Proof of Theorem~\ref{theo:main}}  \label{app:theo:main}

In \cite{Narayanan1991PLP,MinAveCost}, the authors proposed a DA for determining the principal partition sequence (PSP) for a clustering problem. Since the fundamental partition is one of the partitions in PSP \cite{ChanMMI,ChanInfoCluster}, we adapt DA to MDA to just determine the fundamental partition. A similar approach can be found in \cite{ChanInfoCluster}. Based on the studies in \cite{Narayanan1991PLP,ChanInfoCluster}, if the CoordSatCapFus algorithm is able to determine the minimum and the minimal/finest minimizer of the Dilworth truncation problem $\min_{\Pat\in \Pi(V)} \FuHash{\alpha}[\Pat]$ for a given value of $\alpha$, the MDA algorithm outputs $\RACO(V)$, the fundamental partition and an optimal rate $\rv_V \in \RRACO^*(V) = B(\FuHashHat{\RACO(V)},\leq)$. In addition, the value of $\alpha$ of the MDA algorithm converges monotonically upward to $\RACO(V)$.

Now, we show that CoordSatCapFus algorithm determines the finest minimizer of $\min_{\Pat\in \Pi(V)} \FuHash{\alpha}[\Pat]$. For $\FuHash{\alpha}$, Consider the (original/general) CoordSatCap algorithm \cite{Fujishige2005}:
\begin{enumerate}[step 1:]
    \item Initiate $\rv_V$ such that $\rv_V \in P(\FuHash{\alpha},\leq)$;
    \item For each dimension $i \in \Set{1,\dotsc,|V|}$, do $ \rv \leftarrow \rv + \hat{\xi} \chi_{\phi_i}$, where $\hat{\xi}$ is the \text{saturation capacity}
        \begin{equation} \label{eq:SatCap}
            \hat{\xi} = \min\Set{\FuHash{\alpha}(X) - r(X) \colon \phi_i \in X \subseteq V}.
        \end{equation}
\end{enumerate}
$\hat{\xi}$ in \eqref{eq:SatCap} is the maximum increment in $r_{\phi_i}$ such that the resulting $\rv_V$ is still in $P(\FuHash{\alpha},\leq)$, hence the name saturation capacity. Due to the intersecting submodularity of $\FuHash{\alpha}$, \eqref{eq:SatCap} is an SFM problem and the CoordSatCap algorithm finally updates $\rv_V$ to a vector/rate in $B(\FuHashHat{\alpha},\leq)$ with $r(V) = \FuHashHat{\alpha}(V)$.

The minimal minimizer of $\min_{\Pat\in \Pi(V)} \FuHash{\alpha}[\Pat]$ is determined as follows. Let $\hat{X}_{\phi_i}$ be the minimal minimizer of \eqref{eq:SatCap} for dimension $\phi_i$. By iteratively merging dimensions $\phi_i,\phi_j \in V$ such that $\phi_i \in \hat{X}_{\phi_j}$ until there is no such pair left, we can determine the finest partition in $\Pi(V)$ that minimizes $\FuHash{\alpha}[\Pat]$ \cite{Bilxby1985,Fujishige2005,MinAveCost}.\footnote{The minimal minimizer of $\min_{\Pat\in \Pi(V)} \FuHash{\alpha}[\Pat]$ corresponds to the minimal separators of a submodular system with the rank function being $\FuHashHat{\alpha}$. Define the partial order $\preceq$ as $\phi_i \preceq \phi_j$ if $\phi_i \in \hat{X}_{\phi_j}$. Let $G(V,E)$ be the digraph with the edge set constituted by edges $e_{\phi_i,\phi_j} \in E$ if $\phi_i \preceq \phi_j$. The minimal separators are the strongly connected components of the underlying undirected graph of $G(V,E)$. The procedure that updates $\Pat^*$ in Appendix~\ref{app:theo:main} is exactly the one that determines these minimal separators. For more details, we refer the reader to \cite{Bilxby1985,Fujishige2005}.} The implementation is as follows. Initiate $\Pat^*=\Set{\Set{\phi_i} \colon i \in V}$ at the beginning of the CoordSatCap algorithm. After obtaining each $\hat{X}_{\phi_i}$ for $i$ in step 2, do the followings:
\begin{itemize}
    \item find all elements in $\Pat^*$ that intersect with $\hat{X}_{\phi_i}$, i.e., determine $\X = \Set{X \in \Pat^* \colon X \cap \hat{X}_{\phi_i} \neq \emptyset}$;
    \item merge all the elements in $\X$ to form a single element in $\Pat^*$ by $\Pat^* = (\Pat^* \setminus \X) \cup \tilde{\X}$.
\end{itemize}
$\Pat^*$ is the minimal minimizer of $\min_{\Pat\in \Pi(V)} \FuHash{\alpha}[\Pat]$ at the end of the CoordSatCap algorithm. It is easy to see that by letting $\rv_V = (\alpha-H(V))\chi_V$ we have $\rv_V \in P(\FuHash{\alpha},\leq)$ initially. Let $\Phi$ be any linear ordering of $V$. We have
    \begin{multline} \label{eq:PolyRank1}
        \min\Set{ \FuHash{\alpha}(X) - r(X) \colon \phi_i \in X \subseteq V} \\ = \min\Set{ \FuHash{\alpha}(X) - r(X) \colon \phi_i \in X \subseteq V_i}
    \end{multline}
where $V_{i} = \Set{\phi_1,\dotsc,\phi_{i}}$ due to the monotonicity of the entropy function $H$ \cite{Fujishige2005}.\footnote{This property has also been used in \cite{MiloIT2015,CourtIT2014} for solving the non-asymptotic CO problem.}

\begin{lemma} \label{lemma:PolyRank2}
    Let $\Pat^*$ be the partition that is updated in each iteration of the CoordSatCap algorithm as described above,
    \begin{multline}
        \min\Set{\FuHash{\alpha}(X) - r(X) \colon \phi_i \in X \subseteq V}  \\ = \min\Set{\FuHash{\alpha}(\tilde{U}) - r(\tilde{U}) \colon \Set{\phi_i} \in U \subseteq \Pat^*}. \nonumber
    \end{multline}
    Let $\hat{X}_{\phi_i}$ and $U_{\phi_i}^*$ be the minimal minimizer of the LHS and RHS, respectively, of the equation above. Then, $\hat{X}_i=\tilde{U}_i^*$.
\end{lemma}
\begin{IEEEproof}
    For any $X \subseteq V$, let $\Y=\Set{Y \in \Pat \colon Y \cap X \neq \emptyset}$. We have
    \begin{equation}
        \begin{aligned}
            & \quad \FuHash{\alpha}(X) - r(X) - \FuHash{\alpha}(\tilde{\Y}) + r(\tilde{\Y})  \\
            & = \FuHash{\alpha}(X) - \FuHash{\alpha}(\tilde{\Y}) + r( \tilde{\Y} \setminus X )  \\
            & = \FuHash{\alpha}(X) - \FuHash{\alpha}(\tilde{\Y}) + \sum_{Y \in \Y}  \big( \FuHash{\alpha}(Y) - \FuHash{\alpha}(Y \cap X)  \big)  \geq 0,
        \end{aligned} \nonumber
    \end{equation}
    where the last inequality is obtained by applying submodular inequality \eqref{eq:SubMIneq} inductively over intersecting subsets. The minimality of $\tilde{U}_i^*$ over all $X \subseteq V$ such that $\phi_i \in X$ can also be seen by induction. So, $\hat{X}_i=\tilde{U}_i^*$.
\end{IEEEproof}

Based on \eqref{eq:PolyRank1} and Lemma~\ref{lemma:PolyRank2}, we can implement the CoordSatCap algorithm by a fusion method as in the CoordSatCapFus algorithm, where steps 8 and 9 are equivalent to the procedure that updates $\Pat^*$ as described above.

\bibliographystyle{ieeetr}
\bibliography{AsymCO_BIB}

\end{document}